\documentclass[sigconf]{acmart}

\AtBeginDocument{%
  }

\copyrightyear{2026}
\acmYear{2026}
\setcopyright{cc}
\setcctype{by}
\acmConference[CHI '26]{Proceedings of the 2026 CHI Conference on Human Factors in Computing Systems}{April 13--17, 2026}{Barcelona, Spain}
\acmBooktitle{Proceedings of the 2026 CHI Conference on Human Factors in Computing Systems (CHI '26), April 13--17, 2026, Barcelona, Spain}
\acmDOI{10.1145/3772318.3791920}
\acmISBN{979-8-4007-2278-3/2026/04}

\renewenvironment{quote}
  {\list{}{\rightmargin=0.5cm \leftmargin=0.5cm}%
   \item\relax}
  {\endlist}

\usepackage{enumitem}

\usepackage{xcolor,colortbl}
\usepackage{multirow}
\usepackage{array}
\usepackage{siunitx}

\usepackage{xspace}
\usepackage{balance} 
\usepackage{hyperref} 
\usepackage[nameinlink]{cleveref}

\begin{document}

\title[AI Disclosure in Freelance Work]{``Better Ask for Forgiveness than Permission'': Practices and Policies of AI Disclosure in Freelance Work}


\author{Angel Hsing-Chi Hwang}
\affiliation{%
  \institution{University of Southern California}
  \city{Los Angeles}
  \country{United States}
  }
\email{angel.hwang@usc.edu}

\author{Senya Wong}
\affiliation{%
  \institution{University of Southern California}
  \city{Los Angeles}
  \country{United States}
  }
\email{senyawon@usc.edu}

\author{Baixiao Chen}
\affiliation{%
  \institution{Emory University}
  \city{DeKalb County}
  \country{United States}
  }
\email{shawn030131@gmail.com}

\author{Jessica He}
\affiliation{%
  \institution{IBM Research}
  \city{Seattle}
  \country{United States}
  }
\email{jessicahe@ibm.com}

\author{Hyo Jin Do}
\affiliation{%
  \institution{IBM Research}
  \city{Cambridge}
  \country{United States}
  }
\email{dohyojin90@gmail.com}

\renewcommand{\shortauthors}{Hwang et al.}

\begin{abstract}
  The growing use of AI applications among freelance workers is reshaping trust and relationships with clients. This paper investigates how both workers and clients perceive AI use and disclosure in the freelance economy through a three-stage study: interviews with workers and two survey studies with workers and clients. Findings first reveal a key expectation gap around disclosure: Workers often adopt passive disclosure practices, revealing AI use only when asked, as they assume clients can already detect it. Clients, however, are far less confident in recognizing AI-assisted work and prefer proactive disclosure. A second finding highlights the role of unclear or absent client AI policies, which leave workers consistently misinterpreting clients' expectations for AI use and disclosure. Together, these gaps point to the need for clearer guidelines and practices for AI disclosure. Insights extend beyond freelancing, offering implications for trust, accountability, and policy design in other AI-mediated work domains.
\end{abstract}




\maketitle

\section{Introduction}
In March 2023, a freelancer on Upwork posted a question on the platform’s community forum, asking about ``\textit{Upwork’s stance on the use of ChatGPT.}''
What began as a simple request for clarification escalated into heated discussions that lasted nearly a year before the platform shut down the thread. 
By the time we wrote the paper in 2025, Upwork had not yet released a formal, platform-wide guideline on AI use in response to the question.

The Upwork case is not a one-of-a-kind incident. 
Workplaces across industries and domains are facing similar challenges in clarifying what constitutes acceptable and responsible AI use~\cite{kyi_governance_2025}. 
Even in organizations that have issued guidelines for AI use, workers continue to struggle with questions about the appropriate extent of AI assistance and which use cases are considered permissible~\cite{Zirar2023, HR-dive, Forbes-report-AI-confusion}.

While navigating AI use is complex in itself, the challenge of \textit{disclosing} AI use can be more substantial~\cite{cheong_penalizing_2025}. 
Across domains, researchers and practitioners have identified widespread under-the-hood AI use~\cite{zhang_secret_2025}. Such practices pose serious risks, including potential violations of privacy, client contracts, and professional standards — risks that managers, executives, and clients may not fully recognize. 
Furthermore, when such secret use of AI is eventually revealed, it can severely damage trust relationships, in some cases causing irreparable harm.

These dynamics are even more acute in freelance contexts. 
Unlike traditional workplaces, freelancers typically lack long-term contracts, organizational protocols, or clear escalation channels to mediate disputes about AI use~\cite{alvarez_de_la_vega_design_2022}.
Trust between workers and clients\footnote{Throughout this paper, we refer to \textbf{\textit{workers}} as freelancers who seek and perform jobs on freelance platforms, while \textbf{\textit{clients}} are those who hire and work with freelancers.} is fragile: each project depends on maintaining a client’s confidence, and once damaged, the relationship rarely recovers.
As a result, freelancers face especially high stakes when deciding whether, when, and how to disclose AI use.

At the same time, the freelance space also offers unique opportunities to shape practices and policies around AI disclosure. 
Without organizational constraints, clients on freelance platforms have greater direct decision-making power and flexibility. They can communicate expectations, structure “AI policies,” and negotiate agreements with workers in ways that are responsive to specific projects, contexts, and, most importantly, the unique work dynamics with individual workers~\cite{ma_speculative_2025}.

The present work leverages the unique context of freelance work for two reasons. 
First, because individual workers and clients have autonomy to define their own AI use and expectations, studying them offers a rare window into how decisions and interpretations around AI policy are formed, in contrast to corporate settings where policy development processes are often opaque. 
Second, as workers and clients must navigate technological changes and new norms without organizational guidance, studying freelance platforms reveals what kinds of support, infrastructures, or collective strategies might benefit other decentralized platforms, a domain where AI governance becomes increasingly important.

We conducted a three-stage study that iteratively gathered and synthesized perspectives from both workers and clients. 
We began by interviewing workers to profile their disclosure practices (Study~1.1, $N=41$), and then surveyed a larger sample of workers to validate these insights (Study~1.2, $N=100$). 
We mirrored the survey content in a subsequent survey with clients (Study~2, $N=145$), enabling us to compare perspectives between the two groups and identify points of alignment and divergence. Because AI policies consistently emerged as central in shaping disclosure, we also asked clients to share the policies they had developed. 
In Study~3 ($N=100$), we presented these policies back to workers, asking them to interpret the forms of AI use and disclosure they implied. This design allowed us not only to map current practices but also to examine how policies are understood, misinterpreted, and negotiated in freelance contexts.

Through this multi-stage study, we examined how workers approach AI use and disclosure in their freelance work (RQ1), how these practices align—or misalign—with clients’ expectations (RQ2), and how clients’ AI policies shape workers’ interpretations and behaviors (RQ3). In summary, we highlight the following key findings:

\begin{enumerate}[leftmargin=*]
    \item Workers consistently misinterpreted the scope of permitted AI use under clients’ policies. Misunderstandings were most pronounced under policies that encouraged partial use of AI (categorized as “Use AI for minor tasks” or “Use AI for major tasks” in our analyses), as clients and workers often disagreed on what constitutes minor versus major AI use in practice.

    \item We identify five types of AI disclosure practices among freelancers, with \textit{Passive Disclosure} (i.e., workers disclose AI use only when clients explicitly ask) emerging as the most common.
    
    \item Clients generally hold higher expectations for workers disclosing AI use proactively; however, once clients implement explicit AI policies, they often view disclosure as less necessary.
    
    \item Atop (3.), Workers tended to disclose more proactively when encountering policies that encouraged AI use, whereas clients more often demanded disclosure under policies that discouraged or prohibited AI use.
    
    \item Our qualitative analysis of clients’ AI policies revealed several common issues: reliance on vague responsibility clauses, 
    an emphasis on restrictions (particularly around data protection and privacy) rather than clear guidance on appropriate use cases, and directives to avoid AI only at the “final decision stage,” overlooking how earlier steps in the workflow can also critically shape final outputs.
\end{enumerate}

This work contributes to CHI by advancing understanding of how users—specifically freelance workers—adopt, disclose, and negotiate AI use in their tasks. We examine how workers integrate AI into their workflows, how clients respond to and govern these practices, and how disclosure policies are interpreted and enacted in real settings. As formal regulations increasingly mandate AI disclosure, it is essential to understand how users and stakeholders make sense of such guidelines so that policy and regulatory approaches can evolve alongside rapidly advancing AI technologies.

We argue that the freelance space offers a unique window into how individuals form decisions around AI use and AI policies. Although freelancers may differ from workers in corporate environments, the freelance context provides a less opaque setting for observing how AI policies are created, implemented, adhered to, and adapted over time—conditions that are far more difficult to study within organizations. Moreover, because freelance workers and clients operate independently and with limited institutional support, they face greater challenges in navigating AI use on their own. The misalignments we observe between workers’ and clients’ expectations further inform where future AI literacy initiatives could provide meaningful support.

\section{Background and Related Work}

\subsection{Navigating Freelance Workplace in the Age of Generative AI}

As seen in all work settings, the use of generative AI has become increasingly prevalent on freelance platforms~\cite{forbes_freelance2023}. 
Although workers often grapple with the limitations of these tools, most remain eager to adopt AI to streamline their tasks and improve productivity~\cite{dolata_development_2024}. 
Clients’ attitudes toward workers’ AI use, on the other hand, show greater variation and are evolving rapidly. 
Some clients impose strict prohibitions on AI, while others hold what recent research has described as “inflated expectations”~\cite{dolata_development_2024}; namely, clients’ highly optimistic views of AI’s capabilities lead them to set unrealistic demands, assuming that workers can meet them by leveraging AI assistance.

Successful freelance work typically depends on strong client–worker relationships, and these dynamics have only grown more important in the age of generative AI~\cite{huang_design_2024}. 
Effective communication, clear goals, and trust are central to high-quality outcomes~\cite{hsieh_designing_2023, alvarez_de_la_vega_design_2022, hulikal_muralidhar_collaboration_2022}. Positive client–worker relationships also improve retention, reducing the costs of repeatedly searching for new talent and initiating new contracts~\cite{dolata_development_2024}. 
For workers, securing longer-term collaborations—rather than fragmented, one-off tasks—provides greater job stability and supports career development~\cite{hsieh_designing_2023, munoz_platform-mediated_2022}.

As workers increasingly adopt AI to facilitate their tasks, it remains largely unknown whether and how such practices might compromise their long-term, mutually beneficial relationships with clients~\cite{Hwang_llm-authenticity_2024, chen_missing_2025, porquet_copying_2025}. When compromises arise, how can workers and clients mitigate potential harms of AI adoption, and under what conditions might disclosure foster trust rather than erode it~\cite{cheong_penalizing_2025, woodruff_how_2024, schilke2025transparency}? The current research addresses this understudied topic by examining the interplay between workers’ AI use, disclosure practices, and clients’ expectations in freelance contexts.

\subsection{Perceptual Harms and Secret Use of AI}

Perceived AI use, the perception that someone has used AI to produce content or complete work, can make a significant impact in both professional and social contexts~\cite{kadoma_generative_2025, shelby_sociotechnical_2023}. 
In many cases, perceived use of AI causes negative evaluations of both the work itself and the individual who created it~\cite{draxler_ai_2024, jakesch_human_2023}.
This phenomenon has long been observed in research on AI-mediated communication, where messages believed to be AI-generated are judged more harshly, and their authors are likewise viewed as less likable~\cite{jakesch_ai-mediated_2019, hancock_ai-mediated_2020, mieczkowski_ai-mediated_2021}.
Two mechanisms help explain this negative bias. First, AI use is often perceived as less intentional, suggesting that the worker invested less thought and effort in the output~\cite{tigre2023artificial, Heimstad_effort_heuristics_2025, Messer_authenticity_effort_2024}. 
Second, judgments about content quality can be shaped by ``effort heuristics'': people tend to associate greater effort and time with higher quality~\cite{kurosu_bias_2018}.
Because AI-assisted work is assumed to be produced more quickly and with less effort, it is often evaluated as inferior, regardless of the actual quality of the output~\cite{heimstad_machine_2025}.

Moreover, recent research has asked whether the perceptual harms of AI use disproportionately affect certain groups, and whether some individuals are more likely than others to be assumed to rely on AI~\cite{otis2024global,kadoma_generative_2025,gai2025competence}. 
For instance, studies suggest that novices are penalized more heavily when their work is perceived as AI-generated~\cite{Hwang_Yang_AOM_2025}, receiving more negative evaluations than what established professionals encounter~\cite{sarkar2025ai}.
In contrast, experienced workers can sometimes frame AI use as a sign of sophistication, efficiency, or a competitive edge, particularly in domains where technical expertise and productivity are highly valued~\cite{Hwang_llm-authenticity_2024, Hwang_Yang_AOM_2025, Sun_gemini-knowledge-worker_2025, Vaccaro_haic-review-nature_2024}. 
These dynamics highlight that the social meaning of AI use is not uniform, but shaped by the worker’s status, experience, and professional context.

Given these varied reactions to AI use, many workers are acutely concerned about how others will judge them~\cite{reif2025evidence,giray2024ai}. 
Indeed, the perceptual harms of AI use can be especially salient: as highlighted in recent reviews, simply labeling identical content as AI-assisted is enough to affect viewers' perceptions of its credibility and likability~\cite{baek_effect_2024,toff2025or}. 
In response, many workers choose to conceal their AI use~\cite{glynn2024suspected, zhang_secret_2025, nytimes}.
Recent research indicates that such concealment is more prevalent in tasks that are high-stakes, sensitive, or closely tied to personal assessment (e.g., academic writing or self-presentation), as users exhibit heightened concern about negative external judgment regarding their use of AI in these scenarios~\cite{zhang_secret_2025}.
These features closely resemble freelance work, where workers’ reputations hinge on client evaluations. Because public reviews and platform scores strongly influence their ability to secure future opportunities, freelancers may be particularly motivated to conceal their AI use, fearing that disclosure could undermine trust and perceived quality of their work from clients' perspectives.

\subsection{Transparency, Attribution, and Disclosure of AI Use}

Yet in many contexts, disclosure is no longer optional~\cite{he_exploring_2025, he_which_2025}. 
A growing number of transparency regulations—both at the organizational and governmental levels (e.g., the EU AI Act and the California AI Transparency Act)—require users to disclose their use of AI in specific cases. 
As Zhang et al.~\cite{zhang_secret_2025} reviewed, numerous policies across different domains require workers to acknowledge AI assistance.

An unresolved challenge, however, is that these policies rarely specify \textbf{\textit{how}} disclosure should occur or what form it should take. Most adopt a binary framing—either AI was used or not—but this approach fails to capture the varied degrees, contexts, and nuances of AI assistance in practice. Recent research has demonstrated that such binary approaches to attribute AI use are inadequate~\cite{he_exploring_2025, he_which_2025, baek_effect_2024}: they not only obscure the complexity of AI involvement but can also harm workers by inviting negative judgments even when AI was used minimally or responsibly. More granular and context-sensitive approaches to AI attribution are therefore needed to balance transparency with fairness for workers~\cite{dolata_more_2025, kyi_governance_2025, stalnaker_developer_2025}.

We anticipate that the practice of AI disclosure will be particularly challenging in the freelance space, where transparency obligations intersect with the dynamics of client–worker relationships. 
As with many practices in freelancing, decisions about when and how AI use should be disclosed are likely to depend on mutual agreements between individual workers and clients. 
Because each collaboration follows its own trajectory, one-size-fits-all disclosure policies are unlikely to suffice~\cite{zhang_data_2024, hulikal_muralidhar_collaboration_2022}. 
Instead, disclosure arrangements may need to be personalized, recognizing that clients’ expectations and workers’ practices vary across projects, tasks, and domains~\cite{li_bottom-up_2022, hsieh_designing_2023}.

\section{Methods}
We conducted a three-stage study to examine how freelance workers and clients navigate AI use and disclosure. The full protocol was approved by the authors’ Institutional Review Board (IRB). 
Below, we present an overview of our methods; the full methodological details are provided in \Cref{app:method-detail}.
We attach the full questionnaires and interview protocols. 
Additional information about participants demographics are reported in \Cref{app:demographics}. \Cref{app:freelance-type} includes a breakdown of freelance service types that workers offered and clients sought.

\textbf{\textit{Study 1.1 (Interviews with Workers).}}
We interviewed 41 freelancers (46.3\% male, 46.3\% female, age: $40.35 \pm 11.29$) across freelance platforms about their AI use, disclosure practices, and concerns. 45-60-minute long interviews through Zoom were recorded, transcribed and analyzed using open coding and iterative team discussions to identify themes.

\textbf{\textit{Study 1.2 (Survey with Workers).}}
Drawing from Study 1.1, we developed a Qualtrics survey with both closed- and open-ended questions on AI use, disclosure (no, passive, active), and experiences with client policies. $N=100$ freelance workers (53\% male, 39\% female, age: $36.4 \pm 9.8$) participated in Study~1.2. 

\textbf{\textit{Study 2 (Survey with Clients).}}
To mirror worker perspectives, we surveyed 145 clients (gender: 51.38\% male, 33.33\% female; age: $42.31 \pm 11.20$) from freelance platforms about expectations for AI use, preferred disclosure practices, and any “AI policies” they had in place. This enabled direct comparison between workers and clients.

\textbf{\textit{Study 3 (Survey with Workers on Client Policies).}}
Finally, we presented client-defined policies (collected in Study 2) to 100 new workers (gender: 50.00\% male, 48.96\% female, age: $37.29 \pm 9.88$). Each worker evaluated one policy from each of five categories (e.g., “no AI use,” “AI for minor tasks”), rating their interpretations, intended practices, and disclosure strategies.

\section{Findings}

\subsection{AI Use in Freelance Work}

Most freelancers (78.83\%) reported using AI in their work, often describing it as essential for meeting client expectations of speed and acceptable quality.
Workers emphasized that clients cared more about timely deliverables than the process itself, and many felt the urge to adopt AI 
as they would otherwise ``\textit{fall behind}'' and ``\textit{lose my jobs as there are people charging half my rate overseas.}''

Aside from 14.06\% of clients who discouraged AI entirely and 4.69\% who believed AI use was not relevant in their business functions, the majority (81.25\%) of clients encouraged AI use in some capacities. 
Specifically, we synthesized various degrees of AI use first based on interviews and surveys with workers and later confirmed through surveys with clients; these include:
Use AI as much as possible (39.06\%), Use AI for minor tasks (23.44\%), Use AI for major tasks (14.06\%), and
Use AI to automate work entirely (4.69\%).

As we recruited workers and issued job contracts through freelance platforms -- identical to the typical process used by clients -- we noted that these platforms do not provide any tools for detecting AI use. We confirmed this through interviews with workers, ensuring that their perceptions matched this platform reality. Despite the absence of platform-supported detection mechanisms, nearly two-thirds of workers (63.5\%) believed that clients could identify AI use most of the time (75\%) simply by examining the work product. As one participant explained, ``\textit{if I relate to my day job as an instructor, I can tell when my students are using AI. So I assume they (clients) can as well. If they didn’t say anything [about AI use], I assume that’s a silent `yes'.}''
In contrast, clients expressed far lower confidence in their own ability to detect AI involvement, reporting only slightly above-midpoint confidence on a 5-point scale ($M = 3.11$, $SD = 0.97$).

\subsection{Misalignment of Accepted AI Use}

Due to lack of explicit guidance, workers often developed their own strategies and inferred job postings, deadlines, and compensation rates to decide when AI use was appropriate. 
In particular, high-volume, short-turnaround tasks with low payments were read as tacit permission to use AI. 
As illustrated in one participant's example, ``\textit{if they (clients) want 500 of these Excel tasks done by next Wednesday, I think that's a clear green light for AI.}''
Notably, most workers believed clients would likely respond positively or even non-reactively to AI use, analogizing it to standard productivity tools such that ``\textit{there shouldn't be a surprise if you use Microsoft Office at work.}''

\begin{figure}[t!]
    \centering
    \includegraphics[width=0.95\linewidth]{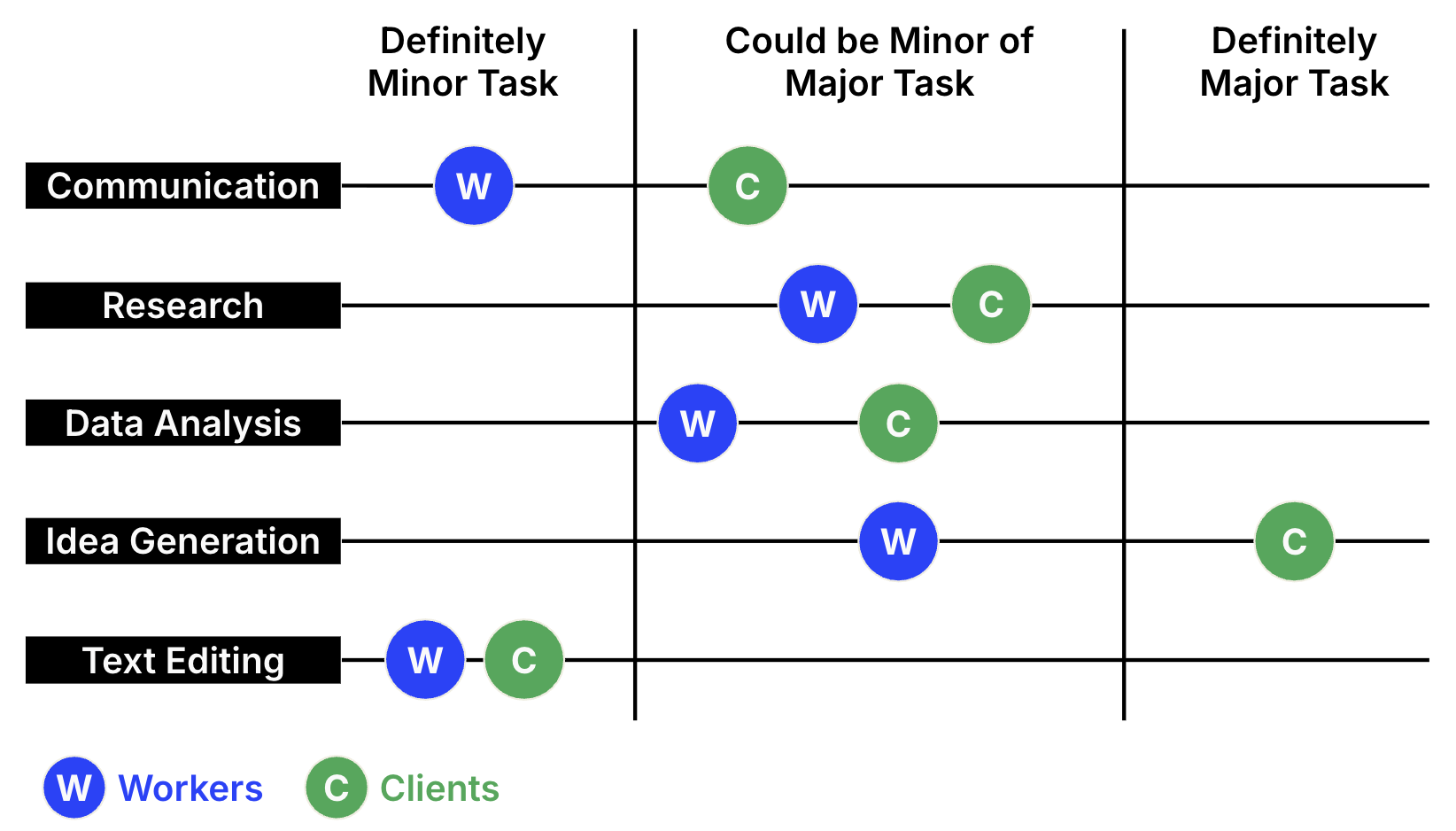}
    \caption{Workers and clients' perceptions of common freelance tasks as major or minor tasks for AI use.}
    \Description{A grid chart compares how workers (blue circles labeled “W”) and clients (green circles labeled “C”) classify different work tasks as minor or major when considering AI use. The chart has three vertical categories: Definitely Minor Task, Could Be Minor or Major Task, and Definitely Major Task. Five tasks are listed on the y-axis: Communication, Research, Data Analysis, Idea Generation, and Text Editing. For each task, one or two circles appear, showing how each group categorizes it: Communication: Workers classify it as definitely a minor task. Clients classify it as possibly minor or major. Research: Workers place it in the possibly minor or major category. Clients place it in the definitely major category. Data Analysis: Workers classify it as possibly minor or major. Clients also classify it as possibly minor or major. Idea Generation: Workers classify it as possibly minor or major. Clients classify it as definitely major. Text Editing: Workers classify it as definitely minor. Clients classify it as definitely minor. A legend at the bottom identifies blue circles as Workers and green circles as Clients. Overall, the figure highlights areas of divergence—especially in Research and Idea Generation—where clients view tasks as more major than workers do, revealing mismatched mental models of what constitutes a “major” task for AI involvement.}
    \label{fig:task-spec}
\end{figure}

Yet workers' assumptions about \textit{how} to use AI for work do not align with clients' actual expectations. As noted above, clients' acceptance depends on the capacity of AI, but workers and clients differ in how they classify ``major'' versus ``minor'' tasks, according to the open-text examples that participants provided through the surveys.
Across responses, some tasks were consistently viewed as minor, others uniformly as major, and some could be interpreted either way. Importantly, the placement of use cases along this minor–major-task spectrum (\Cref{fig:task-spec}) 
diverged between workers and clients.
Below, we highlight the five most frequently mentioned task categories and summarize where perceptions aligned or diverged. 
\begin{itemize}[leftmargin=*]
    \item \textbf{Written Communication.} Workers consistently viewed communication tasks, particularly drafting or preparing emails, as minor. Clients, however, believed that the significance of these tasks depended on their recipients, where customer- or employer-facing communication were viewed as major tasks.
    \item \textbf{Research.} Both groups could view research as either major or minor. AI use for information searches or fact-checking was commonly considered minor. Clients tended to classify research as major when it contributed to important deliverables (e.g., client reports), whereas workers were more likely to see AI-assisted research as minor when it served primarily to offer feedback to their own work.
    \item \textbf{Data analysis.} Workers typically viewed AI-assisted data analysis as minor as long as it did not involve sensitive or proprietary data. Clients, by contrast, tied the classification to the worker’s role: for positions where data analysis is central (e.g., data scientists), it was clearly a major task.
    \item \textbf{Idea generation.} Clients uniformly classified idea generation as a major task. Workers, however, varied: some viewed it as major, while many considered it minor when AI was used only for early-stage brainstorming, inspiration, or final ``touch-ups'' to make outputs appear more novel.
    \item \textbf{Text editing.} Tasks such as proofreading, editing or condensing text, summarization, and spelling or grammar checks were universally considered minor. Workers additionally mentioned drafting text as a minor task, but this use case was observed in clients’ responses.
\end{itemize}

\subsection{Expectation Gaps in AI Disclosure}

\subsubsection{Workers' Approaches to AI Disclosure}
Based on their own notions of accepted use cases, workers developed their own approaches to disclosing AI use.
Synthesizing interview and survey data, we identified five common approaches to AI disclosure, summarized in \Cref{tab:s1-disclosure-persona} (See \Cref{app:method-detail} for details of analysis processes).
The most prevalent was \textit{\textbf{Passive Disclosure}}, in which workers revealed AI use only when clients explicitly asked. 
Importantly, most workers did not withhold disclosure out of fear of negative client reactions. Rather, because AI use was so widespread, proactive disclosure often felt unnecessary or even awkward. 
As one participant explained, ``\textit{it's weird if you have to discuss with your boss which software you use to send emails. So if you do that (AI disclosure), it feels like either your boss is micromanaging or you are being annoying.}''

\begin{table*}[t!]
    \centering
    \resizebox{\textwidth}{!}{%
    \renewcommand{\arraystretch}{1.5}
    \begin{tabular}{p{0.24\textwidth}|p{0.8\textwidth}}
    \toprule
        Type of disclosure practice & Description \\
        \hline
        \textbf{Non-Disclosure or Avoidance\newline n = 3 (7\%)} & 
        A minority of workers stated they would not disclose AI use at all. Reasons included believing it was irrelevant (“it’s just a tool”), fear of clients misinterpreting or undervaluing their work, or concern that disclosure might reduce trust or jeopardize future contracts. \\
        \hline
        \textbf{Passive Disclosure (Disclose only if asked)\newline n = 16, (39\%)} &
        The most common stance is that workers disclose AI use only when clients explicitly ask about it. Many framed this as a passive agreement—AI use does not need to be mentioned unless disclosure is requested. In these cases, workers typically give broad or high-level explanations (e.g., “I used AI to draft ideas” or “I used it as a helper”) without going into technical detail. \\
        \hline
        \textbf{Situational or Conditional Disclosure\newline n = 8, (20\%)} & 
        Workers also described disclosure as context-dependent. For sensitive or competitive projects, they felt disclosure might be more important, while for routine or obvious AI-supported tasks (e.g., transcription, AI-generated voiceovers), disclosure seemed unnecessary. Some reported tailoring the level of detail based on the client’s familiarity with AI or the perceived risks. \\
        \hline
        \textbf{Qualified or Framed Disclosure\newline n = 9, (22\%)} & 
        A sizable group disclosed with disclaimers or framing. They reassured clients that AI was used only as a supportive tool—for brainstorming, proofreading, or saving time—while emphasizing that the final decisions, accuracy checks, and creative control remained with them. This strategy aimed to normalize AI as akin to other everyday tools (e.g., “like a calculator or spell-checker”) while preserving credibility. \\
        \hline
        \textbf{Proactive or Full Transparency\newline n = 5,	(12\%)} &
        Some workers preferred open, upfront disclosure. They described mentioning AI use in proposals, contracts, or notes attached to deliverables. For them, honesty and professionalism required clarifying where AI was involved, especially in tasks like writing, editing, or research. A subset even emphasized “disclosing all parts where AI was used,” sometimes because their employers or clients explicitly required it. \\
    \bottomrule
    \end{tabular}}
    \caption{Five types of workers' AI disclosure practice}
    \label{tab:s1-disclosure-persona}
\end{table*}

Given this passive approach, workers rarely brought up their AI use practices on their own; disclosure conversations were typically client-initiated. Yet when and why clients initiated these discussions varied widely—from clients proactively stating their expectations of AI use before a contract was signed, to raising concerns only when they suspected AI use, to never addressing the issue at all.
Even when AI use was discussed before contracting, clients typically communicated only ``\textit{a general vibe of whether they encouraged or discouraged AI use}'' rather than providing structured or written guidance (e.g., explicit AI-use policies or task-specific expectations). As a result, pre-contract expectations were often informal and ambiguous.

\subsubsection{Mismatched Expectations for AI Disclosure and their Potential Consequences}
\begin{figure}[h!]
    \centering
    \includegraphics[width=0.9\linewidth]{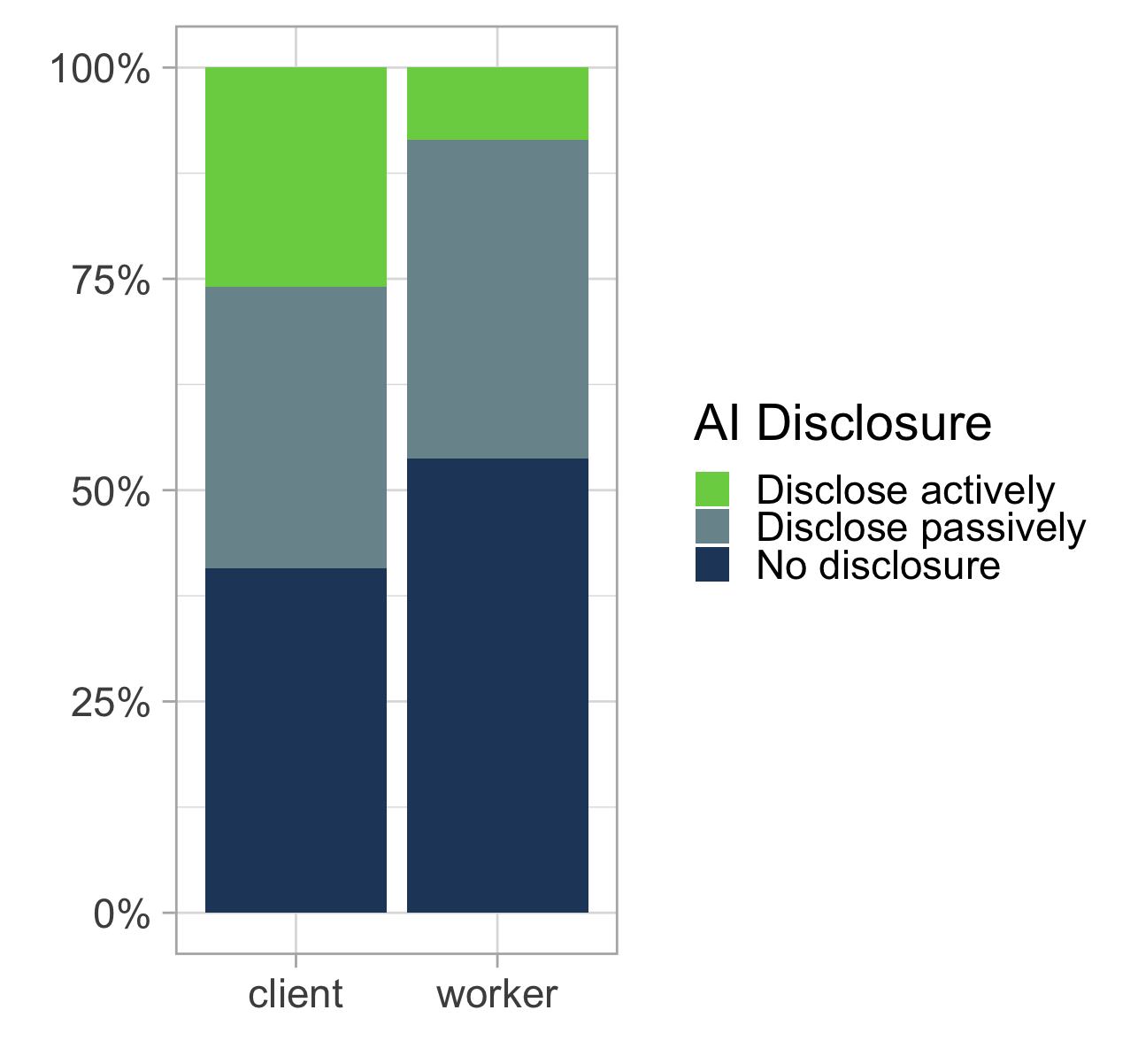}
    \caption{Clients and workers also differed in their attitudes toward disclosure. Significantly more workers than clients believed that disclosure of AI use is not required, whereas significantly more clients expected workers to actively disclose when they use AI.}
    \Description{A stacked bar chart compares clients’ and workers’ AI disclosure practices. Each bar is divided into three categories: Disclose actively (light green, top), Disclose passively (medium gray-green, middle), and No disclosure (dark blue, bottom). For clients, the bar shows roughly one-third “No disclosure,” about one-third “Disclose passively,” and a smaller portion “Disclose actively.” For workers, the bar shows a larger share of “No disclosure,” a substantial share of “Disclose passively,” and a small portion “Disclose actively.” The visualization highlights that workers are less likely than clients to disclose their AI use proactively, with a higher proportion choosing not to disclose at all.}
    \label{fig:ai-disclosure-expectation}
\end{figure}

Under such ambiguities, workers did envision there could be mismatched expectations for AI use between workers and clients.
This matches with the significant differences between clients’ and workers’ expectations around disclosure in survey responses ($t = 2.35$, $p = 0.021$, $d = 0.41$; \Cref{fig:ai-disclosure-expectation}). Over half of workers (53.8\%) reported that disclosure was unnecessary, compared to only 40.7\% of clients. 
Moreover, one quarter of clients (25.9\%) expected workers to proactively disclose AI use, while fewer than one in ten workers (8.6\%) reported that they would disclose in this way. This misalignment highlights an expectation gap in which workers assume disclosure is optional or reactive, while many clients interpret it as a proactive responsibility.

Notably, workers were also conscious about the possible consequences of such misalignment. As with many aspects of freelance platform labor, workers assumed that clients occupied a structurally powerful position when disagreements on AI use and disclosure arise, leaving workers with limited room to negotiate when disputes arose. 
Though none of the workers mentioned direct conflicts with clients over AI use, workers believed clients could terminate a contract or request refunds if they believed AI was used inappropriately. If such cases arise, most workers suggested that they would simply accept the outcome or, at most, file a dispute with limited expectation that the platform would resolve the issue in their favor.

\subsubsection{Evaluating the Tradeoff of AI Disclosure}
Despite these expected consequences, many workers still use AI quietly without proactive disclosure due to their above-mentioned concerns about falling behind and wanting to get more jobs done to yield competitiveness and higher profits on the platforms.
As one participant mentioned, ``\textit{It's all about getting things done. It sucks to put time into a job without earning anything.\footnote{This describes the possible scenario where a client terminates a job contract and requests refunds due to unacceptable AI use \textit{after} a worker has already completed the job.} But it doesn't matter that much once you get your next contract.}''

\begin{figure}[h!]
    \centering
    \includegraphics[width=0.9\linewidth]{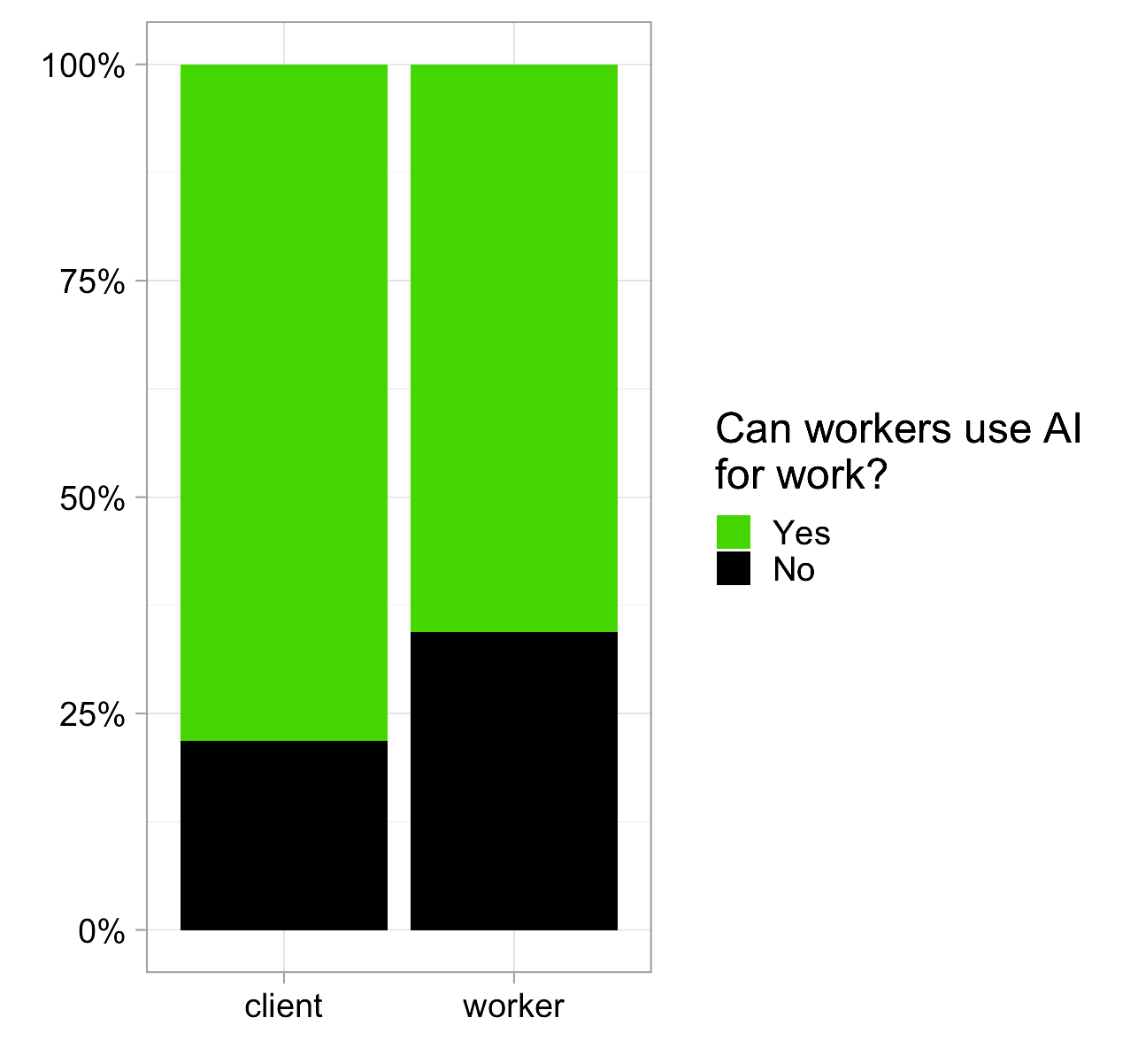}
    \caption{Clients' and workers' responses to whether workers are allowed to use AI for their freelance work. The majority of clients and workers both believe AI use is allowed. The portion of clients allowing AI use is slightly larger than workers' expectations.}
    \Description{Alt-text: A stacked bar chart titled “Can workers use AI for work?” shows two vertical bars: one for clients and one for workers. Each bar is divided into “Yes” (green) and “No” (dark blue). For clients, the bar is approximately 78\% Yes and 22\% No. For workers, the bar is about 66\% Yes and 34\% No. The figure highlights that workers underestimate how often clients permit the use of AI in contract work.}
    \label{fig:ai-yes-no}
\end{figure}

Furthermore, workers acknowledged that while a small minority of clients remained strongly opposed to AI, the likelihood of encountering such clients—and thus facing severe consequences—felt relatively low compared to the substantial benefits of using AI for productivity and competitiveness.
As one participant explicitly stated, ``\textit{it’s better to ask for forgiveness than permission, because the pros of using it (AI) just outweigh the cons so much.}''
As such, even participants working in tech-adjacent roles (e.g., software development, data analysis, technical project management) adopted passive disclosure practices, despite experiencing the most encouraging client attitudes toward AI use.
Moreover, workers may remain underestimating clients’ openness to AI use. While 65.59\% of workers believed they were allowed to use AI for their tasks, 78.13\% of clients said they would permit AI use ($t = 1.74$, $p = .083$, $d = 0.28$; \Cref{fig:ai-yes-no}), suggesting that workers could be using AI without disclosure even when clients accept AI use.



\subsection{(Mis)Interpretations of AI Policies}

Across interviews, workers repeatedly emphasized that they had to make their own judgments about AI use and disclosure because ``\textit{we were never told what to do}'' and ``\textit{I would certainly follow the rules if there were any.}'' Study~3 supports this sentiment: when asked both how they interpreted clients’ AI policies and how they planned to use AI accordingly, workers showed no significant differences between interpretation and planned behavior. In other words, when explicit policies exist, workers intend to follow them and would act based on their interpretations of policies.

In this regard, one might naturally ask whether having AI policies in place would resolve disagreements around AI use and disclosure. 
However, our comparison of clients’ stated policies and expectations in Study~2 with workers’ interpretations in Study~3 reveals that substantial gaps remain: Because workers' interpretations of AI policies constantly mismatched with clients' underlying expectations of these policies, disagreement in AI use and disclosure would likely persist.
Below, we outline these key mismatches.


\begin{figure}[htbp]
    \centering
    \includegraphics[width=0.9\linewidth]{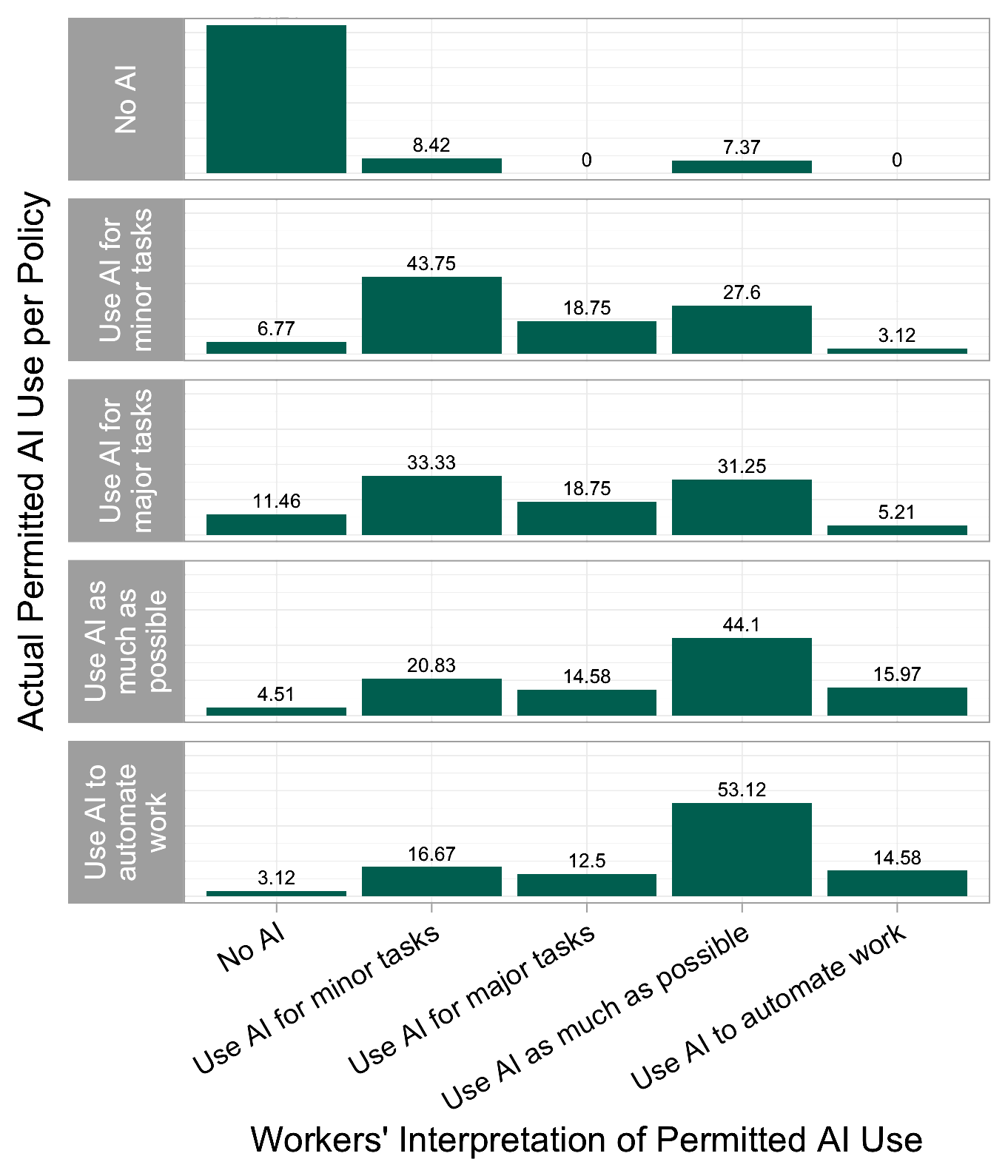}
    \caption{Permitted AI use according to clients' AI policies vs. workers' interpretation of permitted AI use based on each policy.}
    \label{fig:s3-interpret-policy-ai-use-overview}
    \Description{A faceted bar chart compares workers’ interpretations of how much AI use is permitted with the actual AI allowances stated in client policies. The horizontal axis in each panel shows five possible interpretations: No AI, Use AI for minor tasks, Use AI for major tasks, Use AI as much as possible, and Use AI to automate work. Bars represent the proportion of workers who interpreted the policy in each way, with percentages labeled. There are five panels, each representing a different actual policy allowance: First, No AI Allowed: Most workers interpret the policy correctly as “No AI,” with small proportions misinterpreting it as allowing minor tasks or automating work. Second, Use AI for Minor Tasks: Workers’ interpretations spread across categories, with the largest share interpreting it as “Use AI for minor tasks,” and substantial percentages believing major tasks or heavy AI use is also allowed. Use AI for Major Tasks: Workers again show distributed interpretations, with notable shares choosing “minor tasks,” “major tasks,” and “use AI as much as possible.” Third, Use AI as Much as Possible: Workers’ interpretations vary widely, but the largest proportion selects “Use AI as much as possible,” with others selecting adjacent categories. Fifth, Use AI to Automate Work: Many workers correctly interpret the policy as allowing automation, but others interpret it as permitting heavy use or major task support. Overall, the figure highlights pervasive misalignment between what policies actually allow and how workers interpret them, with misinterpretations occurring in every policy category.}
\end{figure}

\begin{table}[htbp]
  \centering
  \resizebox{\linewidth}{!}{
    \begin{tabular}{p{0.55\linewidth}|p{0.1\linewidth}|p{0.1\linewidth}|p{0.1\linewidth}|p{0.2\linewidth}}
    \toprule
         Policy Type & $\beta$ & $S.E.$ & $t$ & $p$ \\
         \hline
         (Intercept) & $0.31$ & $0.11$ & $2.83$ & $0.005^{**}$ \\ 
         Use AI for minor tasks & $0.46$ & $0.12$ & $3.69$ & $<0.001^{***}$ \\
         Use AI for major tasks & $-0.45$ & $0.14$ & $-3.15$ & $0.002^{**}$ \\
         Use AI as much as possible & $-0.84$ & $0.12$ & $-7.20$ & $<0.001^{***}$ \\
         Use AI to automate work & $-1.71$ & $0.14$ & $-11.94$ & $<0.001^{***}$ \\
    \bottomrule
    \end{tabular}}
    \caption{Workers' interpretation of permitted AI use by policy}
    \label{tab:s3-interpreation-gap-lmer}
\end{table}

\begin{figure}[htbp]
    \centering
    \includegraphics[width=0.92\linewidth]{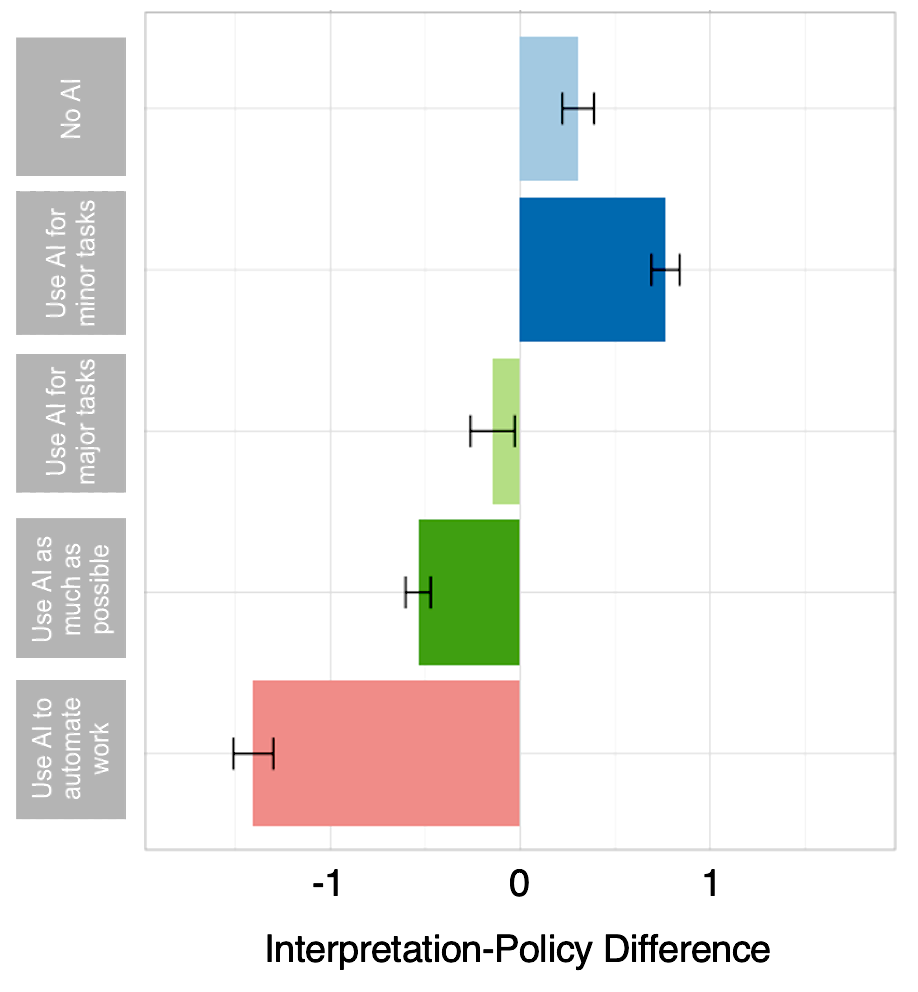}
    \caption{Difference between permitted AI use and workers’ interpretations of AI policies. Positive values indicate that workers believed they were allowed to use AI to a greater extent than the policies actually permitted, while negative values indicate that workers interpreted the policies more restrictively than intended.}
    \label{fig:interpretation-gap-AI-policy}
    \Description{A horizontal bar chart shows the difference between workers’ interpretations of permitted AI use and the actual policy allowances across five policy categories. The x-axis represents the interpretation–policy difference, where positive values indicate workers interpreted the policy as allowing more AI use than intended, and negative values indicate workers interpreted it as allowing less. The y-axis lists the actual policy categories from top to bottom: No AI, Use AI for minor tasks, Use AI for major tasks, Use AI as much as possible, and Use AI to automate work. Each bar shows the mean difference for that category, with a small horizontal black error bar indicating the confidence interval. The pattern indicates: No AI and Use AI for minor tasks both have positive differences, meaning workers tend to overestimate allowed AI use. Use AI for major tasks and Use AI as much as possible are near zero or slightly negative, indicating mixed or slightly conservative interpretations. Use AI to automate work has a large negative difference, meaning workers significantly underestimate how much AI use is allowed. Overall, the figure illustrates systematically different directions of misinterpretation depending on policy strictness.}
\end{figure}

\begin{table}[htbp]
  \centering
  \resizebox{\linewidth}{!}{
    \begin{tabular}{p{0.55\linewidth}|p{0.1\linewidth}|p{0.2\linewidth}|p{0.2\linewidth}}
    \toprule
        Policy Type & $t$ & $p$ & Cohen's $d$ \\
        \hline
        No AI use &  $3.66$ & $<0.001^{***}$ & 0.38 \\
        Use AI for minor tasks & $10.30$ & $<0.001^{***}$ & 0.74 \\
        Use AI for major tasks & $-1.25$ & $0.214$ & $-0.13$ \\
        Use AI as much as possible &  $8.14$ & $<0.001^{***}$ & $-0.48$ \\
        Use AI to fully automate tasks &  $13.35$ & $<0.001^{***}$ & $-1.36$ \\
    \bottomrule
    \end{tabular}}
    \captionof{table}{Interpretation gap by AI policy type}
    \label{tab:s3-interpreation-gap}
\end{table}%

\subsubsection{Permitted AI use.} Workers’ interpretations of client AI policies in Study 3 revealed systematic misalignments. As shown in \Cref{fig:s3-interpret-policy-ai-use-overview}, their judgments about permitted AI use varied widely under nearly all policy types, except when AI was explicitly prohibited. Confusion was particularly pronounced for policies intended to allow AI in major tasks or to support full automation. On average, workers underestimated the extent of permitted AI use compared to what the policies allowed ($t = -2.58$, $p = .010$, $M = -0.17$, $SD = 1.27$). The direction of misinterpretation depended on policy type (\Cref{tab:s3-interpreation-gap-lmer}): when policies discouraged AI (e.g., “No AI” or “AI only for minor tasks”), workers often overestimated what was allowed; when policies encouraged AI (e.g., “major tasks,” “as much as possible,” or “automation”), they consistently underestimated permissions (\Cref{fig:interpretation-gap-AI-policy}).

\subsubsection{Specific AI use cases.} A similar pattern emerged when workers judged specific use cases. Their interpretations diverged significantly from clients’ intentions ($t = 21.56$, $p < .001$, $d = 0.78$), with the largest discrepancies under mid-range policies such as “AI for minor tasks” or “AI for major tasks” (Table~\ref{tab:s3-use-case-lmer}). These were precisely the conditions where partial permission proved most difficult to interpret, leading to frequent misjudgments about whether AI was acceptable for a given task.

\begin{table}[h!]
    \centering
    \resizebox{\linewidth}{!}{
    \begin{tabular}{p{0.55\linewidth}|p{0.08\linewidth}|p{0.08\linewidth}|p{0.08\linewidth}|p{0.18\linewidth}}
    \toprule
         Policy Type  & $\beta$ & $S.E.$ & $t$ & $p$ \\
         \hline
         Baseline: Use AI for minor tasks & & & & \\
         (Intercept) & $0.69$ & $0.06$ & $12.03$ & $<0.001^{***}$ \\ 
         Use AI for major tasks & $0.12$ & $0.09$ & $1.31$ & $0.189$ \\
         Use AI as much as possible & $-0.21$ & $0.07$ & $-3.08$ & $0.002^{**}$ \\
         Use AI to automate work & $-0.40$ & $0.09$ & $-4.32$ & $<0.001^{***}$ \\
    \bottomrule
    \end{tabular}}
    \caption{Workers' interpretation of specific AI use case by policy}
    \label{tab:s3-use-case-lmer}
\end{table}

\subsubsection{Requirement for AI disclosure.} 
\begin{figure}[ht]
    \centering
    \includegraphics[width=\linewidth]{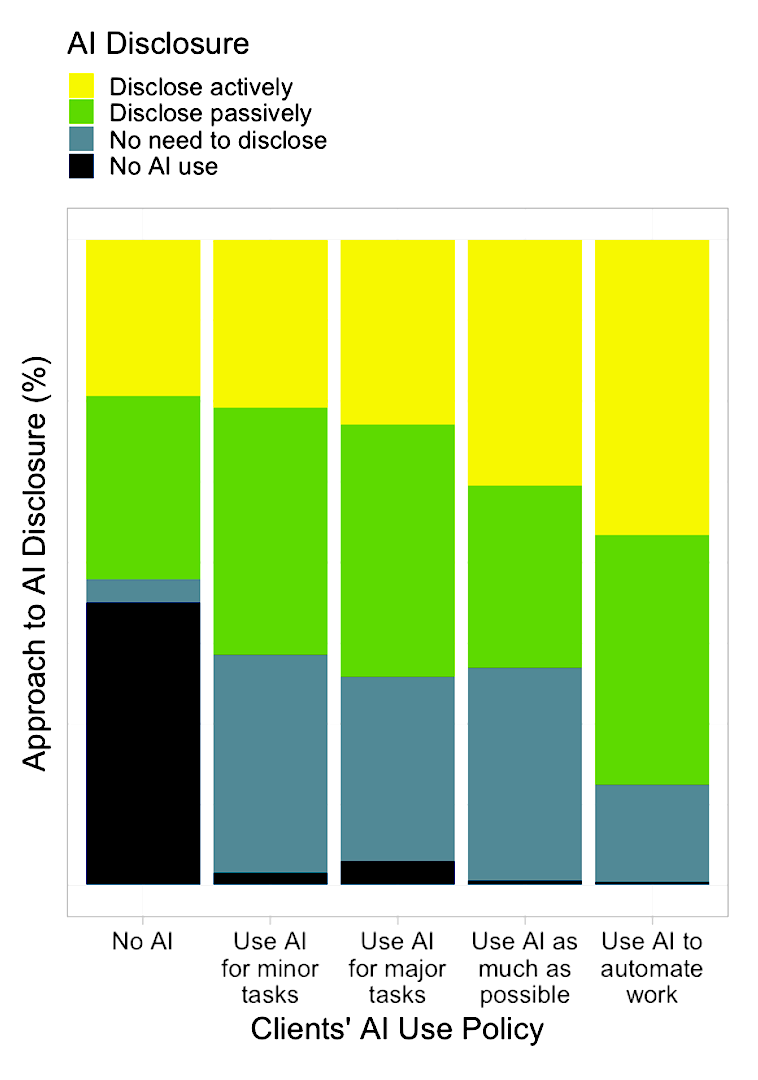}
    \caption{Workers’ interpretations of clients’ disclosure requirements under different AI policies. Workers were more likely to report proactive disclosure of AI use when policies encouraged AI use. 
    }
    \label{fig:AI-policy-worker-disclosure}
    \Description{A stacked bar chart shows how workers’ approaches to AI disclosure vary across five types of client AI-use policies. The x-axis lists the policy categories: No AI, Use AI for minor tasks, Use AI for major tasks, Use AI as much as possible, and Use AI to automate work. The y-axis represents the proportion of workers (in percent) within each policy category. Each bar is divided into four colors: Yellow = Disclose actively. Green = Disclose passively. Teal/blue = No need to disclose. Black = No AI use. Patterns across the bars show that: Under No AI policies, a large proportion of workers report not using AI (black) or needing no disclosure (teal), while smaller shares report passive or active disclosure. As policies become more permissive (minor to major tasks to heavy AI use), passive and active disclosure both increase, and the share reporting no AI use decreases to very small levels. Under the most permissive policies (Use AI to automate work), the majority of workers disclose—either actively or passively—with very little reporting “no AI use.” Overall, the figure highlights that more permissive client policies correlate with greater worker openness to AI disclosure, especially active disclosure.}
\end{figure}

\begin{figure}[htbp]
    \centering
    \includegraphics[width=\linewidth]{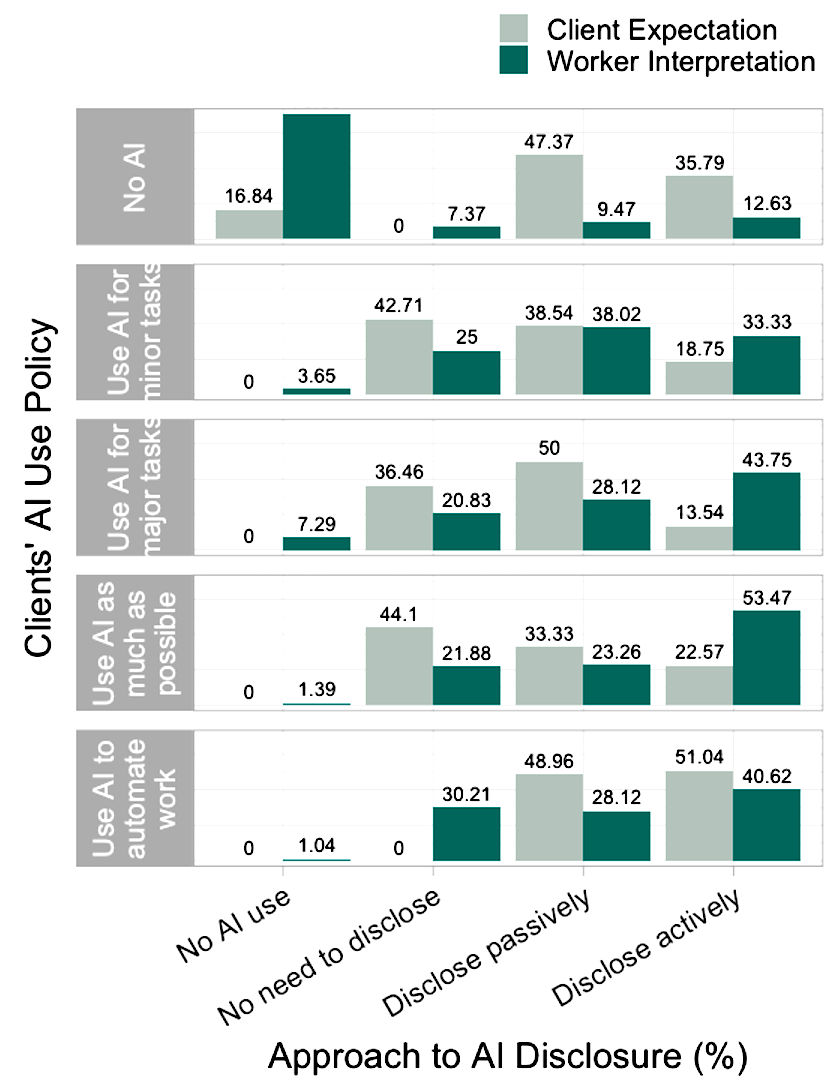}
    \caption{Comparison of clients’ expectations and workers’ interpretations of AI disclosure under different policies. Workers frequently misjudged situations where clients deemed disclosure unnecessary, and consistently overestimated the extent to which clients expected proactive disclosure of AI use.}
    \label{fig:AI-disclosure-client-worker}
    \Description{A multi-panel bar chart compares clients’ expectations for AI disclosure with workers’ interpretations of those expectations across five types of AI-use policies. Each panel represents one policy category along the y-axis: No AI, Use AI for minor tasks, Use AI for major tasks, Use AI as much as possible, and Use AI to automate work. Within each panel, pairs of bars appear for four disclosure approaches shown on the x-axis: No AI use, No need to disclose, Disclose passively, and Disclose actively. Light gray bars represent client expectations. Dark green bars represent worker interpretations. Percentages are labeled above each bar. Key visual patterns: Under No AI policies, clients expect very little AI use or disclosure, yet workers frequently interpret the policy as requiring passive or even active disclosure. For minor task and major task policies, clients expect mostly passive disclosure, but workers are more split across passive and active disclosure. Under more permissive policies (Use AI as much as possible and automate work), client expectations shift strongly toward active disclosure, and workers broadly mirror this—though workers still overestimate the need to disclose compared to clients in several categories. Overall, the figure highlights systematic mismatches between what clients expect and what workers think clients expect, especially around when disclosure is required and how proactively it should be done.}
\end{figure}

We also examined how workers understood disclosure requirements under each policy. As policies became more supportive of AI use, workers reported a greater willingness to disclose (\Cref{tab:s3-disclosure-lmer}, \Cref{fig:AI-policy-worker-disclosure}). Yet these disclosure assumptions did not fully align with client expectations (\Cref{tab:s3-client-worker-disclosure}). Under permissive policies such as “minor tasks” or “major tasks,” most clients indicated that disclosure was unnecessary, while roughly half of workers believed they would still need to proactively disclose. Conversely, when policies prohibited AI, clients expected some form of disclosure in the event of violation, while workers often interpreted prohibition as a reason to remain silent, reasoning that disclosure would only confirm misconduct (\Cref{fig:AI-disclosure-client-worker}).
Finally, we found that workers’ prior experiences shaped their accuracy. General beliefs about whether clients permitted AI or could detect AI use showed no effect. However, prior exposure to explicit AI policies from current or former clients significantly improved workers’ ability to correctly interpret new policies on AI use ($\beta = 0.51$, $S.E. = 0.19$, $t = 2.66$, $p < 0.001$) and AI disclosure ($\beta = 0.69$, $S.E. = 0.22$, $t = 3.20$, $p = 0.001$).

\subsection{A Closer Look at Clients' AI Policies}
To better understand why workers’ interpretations diverged from clients’ policies, we qualitatively analyzed the policies themselves. A first issue was inconsistency in defining “No AI use.” While some employers clearly prohibited all AI involvement, others still allowed limited use for administrative or drafting tasks, making it unclear whether “no AI” meant outright prohibition or partial restriction.

Policies also relied heavily on vague or underspecified language, common examples include ``using AI with common sense'' or ensuring “human judgment.” 
These ambiguous phrases offered little guidance for practical application. Similarly, many policies prohibit AI use on “sensitive topics,” while they rarely define what counts as sensitive, leaving workers to draw their own boundaries.

\begin{table}[ht]
  \centering
  \resizebox{\linewidth}{!}{
    \begin{tabular}{p{0.55\linewidth}|p{0.1\linewidth}|p{0.1\linewidth}|p{0.1\linewidth}|p{0.2\linewidth}}
    \toprule
         Policy Type & $\beta$ & $S.E.$ & $t$ & $p$ \\
         \hline
         Baseline: No AI use & & & & \\
         (Intercept) & $0.64$ & $0.09$ & $6.93$ & $<0.001^{***}$ \\ 
         Use AI for minor tasks & $1.37$ & $0.11$ & $12.59$ & $<0.001^{***}$ \\ 
         Use AI for major tasks & $1.44$ & $0.13$ & $11.49$ & $<0.001^{***}$ \\ 
         Use AI as much as possible & $1.65$ & $0.10$ & $16.05$ & $<0.001^{***}$ \\ 
         Use AI to automate work & $1.44$ & $0.13$ & $11.49$ & $<0.001^{***}$ \\
    \bottomrule
    \end{tabular}}
    \caption{Workers' interpretation of required AI disclosure by policy type}
    \label{tab:s3-disclosure-lmer}
\end{table}%

\begin{table}[htbp]
  \centering
  \resizebox{\linewidth}{!}{
    \begin{tabular}{p{0.55\linewidth}|p{0.1\linewidth}|p{0.2\linewidth}|p{0.2\linewidth}}
    \toprule
         Clients' expectation for AI disclosure & $t$ & $p$ & Cohen's $d$ \\
         \hline
         No AI use &  $2.53$ & $0.022^{*}$ & 0.63 \\
         No need to disclose & $15.01$ & $<0.001^{***}$ & 0.96 \\
         Disclose passively & $1.48$ & $0.140$ & $0.08$ \\
         Disclose actively &  $-12.21$ & $<0.001^{***}$ & $-0.87$ \\
    \bottomrule
    \end{tabular}}
    \caption{Difference between clients' expectations vs. workers' interpretation of AI disclosure.}
    \label{tab:s3-client-worker-disclosure}
\end{table}

Responsibility clauses were another common yet imprecise feature. Nearly all policies referenced worker “responsibilities,” but these statements focused on procedural steps (e.g., reporting use to managers, seeking approval, or verifying outputs), rather than clarifying actual permitted or prohibited AI use cases. 
Furthermore, policies with nearly identical wording often reflected different underlying priorities in terms of confidentiality, accuracy, or organizational control.

Many clients chose to highlight AI’s limitations as a part of their AI policies without offering actionable guidance. 
For example, several policies highlight AI errors, and thus workers should refrain from using AI for "making final decisions." 
This left unclear what kinds of work decisions counted and ignored the various steps workers perform before reaching their final deliverables. 
Likewise, restrictions on logging into AI with company accounts set procedural boundaries but did not clarify which substantive uses were acceptable.

Finally, disclosure requirements revealed tensions. In some policies, disclosure was framed as enabling responsible AI use, yet employers demanded reporting of even routine or minor uses without specifying thresholds. Workers were often told to report AI use to managers but were given little direction on what needed to be disclosed. Notably, many of the employers with such expectations did not have formal policies in place, amplifying the uncertainty.
Taken together, these inconsistencies, ambiguities, and overbroad requirements explain why workers systematically misinterpret client AI policies.

\section{Discussion}
Our multi-stage study examined how freelance workers use and disclose AI, how these practices align—or more often misalign—with clients’ expectations, and how client policies shape interpretation and behavior. We identified five disclosure strategies among workers, with Passive Disclosure, revealing AI use only when explicitly asked, emerging as the most common. Clients, by contrast, generally expected more active disclosure, though this expectation decreased when formal AI policies were in place. Workers also misinterpreted the scope of permitted AI use, particularly under policies that encouraged some but not all forms of AI use. Similarly, workers often overestimated disclosure obligations under permissive policies while underestimating disclosure under prohibitive ones. Finally, our analysis of client AI policies revealed recurring issues: vague responsibility clauses, prohibitions without actionable allowances, and oversimplified stage-based distinctions that failed to reflect the realities of freelance workflows.

Together, these findings underscore a persistent pattern of misalignment: workers’ disclosure practices, clients’ expectations, and formal policies rarely align in straightforward ways. This creates confusion for workers, undermines trust, and exposes the limitations of binary disclosure frameworks in capturing the nuanced ways AI is integrated into freelance work.

\subsection{Misalignment as a Design Problem}
We propose that understanding and shaping norms should be a priority for platforms seeking to resolve misalignments between workers’ and clients’ expectations for AI use and disclosure. Much of the gap we observed stems from the rapid pace of AI adoption, which has led workers and clients to develop different normative assumptions. Workers, for example, increasingly view reporting all AI use as unnecessary and often interpret signals in job descriptions as tacit approval, even when clients did not intend them as such. At the same time, our findings suggest that cultivating certain norms, such as creating an atmosphere that encourages AI use, can make workers more willing to disclose, thereby aligning their practices more closely with what clients hope to see.



Like other forms of collaboration on freelance platforms, AI use and disclosure should be treated as a shared process rather than a one-sided obligation~\cite{hsieh_designing_2023, alvarez_de_la_vega_design_2022, hulikal_muralidhar_collaboration_2022}. Workers were most willing to disclose when clients explicitly encouraged AI use, as such encouragement reduced the stigma they associated with AI---namely, fears of being perceived as unskilled or less dedicated to one's work~\cite{Heimstad_effort_heuristics_2025, Messer_authenticity_effort_2024}. 


To help address these stigma, platforms could help shift these norms by offering clearer guidance and modeling expectations for both parties. For example, platforms might provide policy templates and disclosure scripts that specify which kinds of AI use are acceptable—removing ambiguity and reducing the stigma workers feel when initiating these conversations. They could also highlight positive examples of transparent AI use to normalize disclosure as a routine part of collaborative work rather than a punitive compliance check. By shaping the relational and normative environment in these ways, platforms can make AI transparency more feasible and less risky for both workers and clients.


\subsection{Designing AI Policies in Freelance and Beyond}

Our analysis suggests that most policy failures are fundamentally usability failures: workers struggled not because they resisted disclosure or compliance, but because policies were ambiguous, binary, or disconnected from concrete AI use cases. For example, many “No AI use” policies contradicted themselves—sometimes listing acceptable uses despite an outright ban—creating confusion and encouraging secrecy rather than compliance. A more effective approach is to replace blanket prohibitions with \textbf{\textit{stage-aware allowances}} that specify which practices are acceptable at different points in the workflow (e.g., AI permitted for brainstorming or outlining but restricted for final deliverables).

Policies also need clear examples of what to disclose and how to disclose it, not just restrictions. Research on AI-generated content labeling~\cite{Epstein_ai-label_2023} and platform guidelines (e.g., YouTube’s “Appropriate vs. Inappropriate” lists) show that disclosures are more effective when they provide concrete, operational categories rather than abstract labels. Our findings indicate a similar demand: workers repeatedly asked for examples that clarified which tasks count as “minor” versus “major” AI use, what types of assistance require disclosure, and what a “proper” disclosure statement looks like. For instance, a policy might specify the following in parallel:
\begin{itemize}[leftmargin=*]
    \item Permitted uses: “AI may be used for proofreading, grammar checks, or summarizing client-provided documents.”
    \item Prohibited uses: “AI may not draft or substantially rewrite client-facing materials, generate original designs, or make decisions on budget or strategy.”
    \item Disclosure content: “If AI contributed to outlining or drafting, state which sections were AI-assisted and describe the human review process.” Specifically, this should allow users to show the extent to which content is AI-generated. 
    \item Conditions and rationales of AI use. ``AI is used for image generation for a pitch deck, because the task is not the major job function of the user.''
\end{itemize}
We recommend that future work and users refer to existing AI attribution tools (e.g., IBM Attribution toolkit\footnote{https://aiattribution.github.io/})  to identify relevant items for disclosure. Providing both examples and counterexamples gives workers far clearer guidance than vague phrases such as “avoid sensitive topics” or “use judgment.” Without such specificity, many workers in our study defaulted to borrowing policies from prior employers—policies that often did not match clients’ expectations or freelance workflows, further contributing to mismatch.

Disclosure frameworks should also move beyond binary yes/no categories. AI use exists on a spectrum—from minor proofreading to substantial content generation—and collapsing this spectrum into a single checkbox leads to both over- and under-reporting. A proportional model—what we call a \textbf{\textit{disclosure ladder}}—could tie reporting requirements to the materiality of AI involvement. Minor uses might require no disclosure; supportive uses could warrant a brief note; substantive contributions might require a process summary; and fully automated work could trigger pre-approval or auditing. Explicit thresholds (e.g., disclosure required if more than a set proportion of the deliverable originates from AI) would further reduce guesswork and standardize expectations.

Similarly, policies addressing data privacy must provide actionable guidance rather than broad warnings. Workers need to know which tools are allowed, what types of data may be entered, and how to handle client information across common scenarios. Instead of static documents, platforms could offer scenario cards that illustrate how policies apply to typical tasks; e.g., “Drafting an email with sensitive information,” “Uploading client data to AI tools,” or “Fact-checking using public sources.” Platforms could also support negotiation by allowing clients to specify AI allowances directly in job postings and by attaching automatically generated summaries of relevant rules to contracts.

Finally, to encourage honest disclosure, platforms should build safe harbors and appeals into their systems. Workers in our study repeatedly feared punitive consequences even for minor or permissible AI use. Treating policies as versioned, iterative documents—with change logs, usability testing, and opportunities for clarification—would make them more transparent and trustworthy. These design choices would help shift AI disclosure from a punitive compliance task toward a shared, collaborative norm.

\subsection{Supporting Freelance Workers on Decentralized Platforms}
Our work demonstrates the value of studying freelancers; although freelance workspace differs substantially from conventional corporate settings, studying freelancers enables us to observe the micro-mechanisms through which workers interpret ambiguous AI guidance, how clients translate expectations into policy language, and where interpretive mismatches arise.
Many of these underlying rationales, and the ways they succeed or break down when formalized as policies, are difficult to observe in corporate environments, where AI policy development is typically top-down and opaque.

At the same time, freelancing's structural decentralization makes it more difficult for AI-use norms or disclosure practices to develop collectively.
These challenges reflect on our findings---specifically, constant mismatches between workers and clients' expectations---and resonate with prior research that suggests platform workers repeatedly navigating ambiguity around technological change and shifting norms in the absence of organizational guidance~\cite{jarrahi2019ai, Jarrahi_2021, Sutherland_Jarrahi_Dunn_Nelson_2020}, while AI use only makes navigating these dynamics more challenging~\cite{huang_design_2024,kadoma_generative_2025,Hwang_Yang_AOM_2025,cheong_penalizing_2025}. 

Our findings highlight why platforms must take a more active role in supporting stakeholders. First, mutual misunderstandings are pervasive: clients struggle to formulate clear policies, and workers struggle to interpret them. Second, when disputes arise, workers consistently report that platforms default to client-favoring resolutions, reinforcing power asymmetries long noted in the platform-labor literature~\cite{dolata_development_2024, hsieh_designing_2023, munoz_platform-mediated_2022}. Given that both groups must navigate generative AI without organizational scaffolding, platforms are uniquely positioned to provide infrastructural support. 
We propose two possible directions:

\paragraph{Shared literacy and transparent case resources.}
Platforms and researchers should invest in assembling illustrative cases that are accessible to the public (akin to dark-pattern repositories~\cite{gray-dark-pattern}). This can entail hosting an online library of "bad AI disclosure policies" or collect examples of AI use cases in freelance work to help stakeholders calibrate what constitutes “minor,” “major,” or prohibited AI use and reduce interpretive ambiguity.

\paragraph{Participatory, multi-stakeholder policy-formation systems.}
Recent HCI work demonstrates scalable deliberative systems for producing consensus-driven policy guidelines through structured dialogue, ranking, and expert iteration~\cite{konya2023_democratic-policy-development, kuo-policycraft-2025}. Adapting these methods to freelance ecosystems would allow workers and clients to collaboratively shape AI policies, enabling platforms to implement clearer, democratically informed norms rather than relying solely on client-initiated rules.

\subsection{Implications for AI Regulation Literacy}
We propose that raising awareness of the fast-changing norms around AI use and disclosure should be a core component of AI regulation literacy. Importantly, these norms are shifting in divergent ways for workers and clients. Workers, for instance, need to recognize that even though AI use has become widespread, a client’s statement of “No AI use” may still entail an expectation of strict or minimal usage. Conversely, clients should understand that workers’ tendency toward passive disclosure often reflects an effort to maintain positive work dynamics—avoiding unnecessary micromanagement—rather than an attempt to conceal their practices.


Improving regulation literacy has implications beyond freelance marketplaces. As formal regulations such as the EU AI Act and California’s Transparency in AI Act come into effect, workers across industries will face similar interpretive challenges. Freelance contexts show how regulations are understood on the ground in the absence of institutional supports, offering lessons for policymakers and organizations about how to design training, templates, and communication strategies that build confidence and clarity rather than confusion.

\subsection{Limitations and Future Work}
This study has limitations. Our focus on freelance platforms highlights flexible, project-based relationships that may not generalize to more institutionalized work arrangements. Our sample, though diverse, was shaped by recruitment through specific platforms and may not capture all regional or sectoral practices.

Future research could extend these findings by comparing disclosure dynamics in other precarious labor markets such as gig work or creator economies, and contrasting them with more stable workplaces. Experimental studies might test proportional versus binary disclosure frameworks to assess which mechanisms best foster trust and compliance, while longitudinal designs could track how norms evolve as regulations mature and platform-level policies institutionalize best practices. Together, these directions can deepen understanding of AI governance across diverse work contexts and inform more effective, user-centered approaches to policy design.

\section{Conclusion}
Across interviews and surveys with both workers and clients, our study demonstrates that the governance of AI in freelance work is not simply a matter of drafting rules, but of designing policies, norms, and platform mechanisms that are interpretable and actionable in practice. Misalignments between workers’ disclosure practices, clients’ expectations, and formal policies highlight the limits of binary frameworks and underscore the importance of proportional, stage-aware, and scenario-driven approaches. Freelance marketplaces, as flexible and high-stakes environments, make visible the challenges of operationalizing regulation and offer a valuable testbed for developing usable governance models. By centering usability, negotiation, and trust, future policies can better align the realities of work with emerging regulatory requirements, advancing both effective AI governance and fairer conditions for workers.

\begin{acks}
We thank the participants of the \href{https://datasociety.net/announcements/2025/01/08/what-is-work-worth/}{CHIWORK 2025 Workshop on Navigating Generative AI Disclosure, Ownership, and Accountability in Co-Creative Domains} and \href{https://datasociety.net/announcements/2025/01/08/what-is-work-worth/}{Data\& Society’s Workshop on What Is Work Worth?} for their thoughtful perspectives and feedback, which helped shape this work.
We thank the freelance workers and clients who participated in our study and shared their experiences, without whom this work would not have been possible.
\end{acks}


\bibliographystyle{ACM-Reference-Format}
\bibliography{ref,freelance}

\appendix

\section*{Appendix}

\section{Methodological Details}
\label{app:method-detail}

We conducted a three-stage study, iteratively synthesizing responses from both freelance workers and clients. 
We began with semi-structured interviews with workers to understand their use of AI applications for work, their current approaches to AI disclosure, and any outstanding concerns (Study~1.1).
We then replicated and expanded emerging findings from the interviews through a survey study with a larger group of freelance workers (Study~1.2).
In Study~2, we sought clients' perspectives and expectations on workers' AI use and disclosure through a survey. We designed the survey by mirroring questions and insights that we attained from Study~1.1 and Study~1.2 in order to make direct comparisons between clients' and workers' responses.
We also asked for clients' current ``AI policies'' on whether and how workers should apply and disclose AI at work.
Finally, we presented these AI policies to workers and sought their responses through another round of survey (Study~3).
The full study protocol was reviewed and approved by the authors' Institutional Review Board (IRB).

\subsection{Study 1.1: Interview with Freelance Workers}

\subsubsection{Recruitment Approaches \& Interview Participants}
We recruited $N=41$ freelancers from different industries across different freelance platforms to participate in Study 1.1's interviews.
To diversify the types of freelancers recruited, we posted the same public recruitment message in multiple channels across five freelance platforms (Upwork, Fiverr, Freelancer.com, People Per Hour, and Toptal) and let interested freelancers sign up to participate in the study organically.
Below is the recruitment message:
\begin{quote}
    \textit{Looking for short-term research participants: Share your experience as freelancer!} \\
    \textit{We are a team of researchers at [Anonymous University]. We invite freelancers to share their experiences working and seeking opportunities on freelance platforms.} \\
    \textit{You will participate in a short interview (30~40 minutes) where one of the researchers on our team will ask you questions about your freelance experiences. We will schedule a time for you to participate in the interview at your convenience. We will provide a video-conferencing link for the study session.} \\ 
    \textit{Your contributions will help us understand the challenges and successes faced by freelancers and will inform how we design technologies to address these challenges. This is an opportunity to provide valuable feedback to advance research in this topic area. If you have unique experiences or perspectives, we want to hear from you!}
\end{quote}

\noindent We screened each candidate's profile to ensure they indeed had prior freelance and industry experience and distributed an informed consent form to qualified participants.
We report their demographics, freelance experience, and professional domains in \Cref{app:demographics}.

\subsubsection{Interview Protocol}
Interviews with worker participants consist of three major topics, including (1) understanding workers' freelance experience and background, (2) understanding workers' AI use: their motivations, concerns, and practice of using AI for work, and (3) understanding workers' practice of AI disclosure: their decisions and rationale for disclosing (or not disclosing) AI use as well as their beliefs about the potential impact of such practice.
All interviews were conducted online via Zoom. 
Each interview lasts around 45~minutes.
Below, we enclose the interview protocol:

\textbf{\textit{Introduction:}} Please briefly describe your job and experience as a freelancer.
\begin{itemize}
    \item How long have you been working as a freelancer? At what capacity (e.g., full-time or part-time, how many hours per week)?
    \item What types of work do you do as a freelance worker?
\end{itemize}

\textbf{\textit{AI Use:}} Do you use AI for any part of your work?
\begin{itemize}
    \item If yes, how and what do you use AI for?
    \item How does the use of AI influence your work performance?
\end{itemize}

\textbf{\textit{AI Disclosure:}} Whether and how do you plan to disclose your use of AI for work?
\begin{itemize}
    \item (If you are not using AI at work yourself, do you see the need for people to disclose their use of AI for work to their clients?)
    \item Please elaborate further: What encourages/discourages you from disclosing your use of AI for work?
    \item Do you hold any concern about disclosing your use of AI for work?
    \begin{itemize}
        \item Do you think disclosing the use of AI influences the types of clients who want to work with you and the types of opportunities you could cultivate on the platform? How so?
        \item Do you think disclosing the use of AI influences your work relationship and trust-building with clients? How so?
        \item Do you think disclosing the use of AI influences how you are treated and compensated on the platform? How so?
    \end{itemize}
\end{itemize}

\subsubsection{Data Collection \& Analysis}
All interview sessions were video-recorded through Zoom. 
Researchers also took notes during each interview session, recording participants' responses to our research questions, resulting in a total of 87 pages of notes. We used Otter.ai to transcribe all recorded interviews for data analysis.

We adopted an open-coding approach to data analysis and did not use a predefined codebook to guide the coding process.
We used a digital whiteboard (Miro) to facilitate the coding process. 
For each participant, we put down short summaries of their responses (with corresponding quotes) to each of our research questions on digital sticky notes. For example, within each participant's responses to their thoughts on AI disclosure, we extracted (a) their current practice, (b) their motivation to adopt such a practice, and (c) their expected consequences of AI disclosure and put them down on three separate sticky notes.
For each research question, we then grouped participants' responses by similarity.
The lead author conducted an initial round of coding to produce preliminary clusters of notes for each research question.
The research team then met weekly/bi-weekly to collaboratively review and discuss these notes, iteratively identify and form a working consensus on emerging patterns, themes, and categories across interviews. 
Through this inductive process and team-based reflection, we produce the five categories of AI disclosure practice in \Cref{tab:s1-disclosure-persona} and other themes in the findings.

To reduce subjectivity during data analysis, we also actively sought feedback from fellow researchers outside our team throughout the data analysis process.
Still, we acknowledge that our own backgrounds and experiences can influence analyses~\cite{Adamu-subjectivity-coding}. 
Hence, we share our positionality statement here for transparency:

We formed our research team with three HCI researchers (one with primary training in computer science, two with primary training in social and behavioral science), one AI/machine learning researcher, and one quantitative social science researcher.
All researchers have both industry and academic research experience.
Per our self-identified demographics, the team consists of four Asian females and one Asian male.
We expect the interdisciplinary backgrounds of our research team to contribute to more diverse perspectives throughout the research process.

\subsection{Study 1.2: Survey with Freelance Workers}

\subsubsection{Questionnaire Development}
We constructed a questionnaire by adapting interview prompts into survey items and adding follow-up questions based on emerging findings. For example, when asking how participants disclose their AI use, we included both a free-text response and a multiple-choice item with three options derived from Study~1.1 interviews: no disclosure, passive disclosure (e.g., disclosing only when clients ask), and active disclosure (e.g., proactively reporting how AI is used to clients). Because several interviewees also noted that their AI use depends on client policies, we added questions (Items 7–9) on whether participants had encountered such guidelines and how they responded. The survey was administered via Qualtrics, took approximately 30 minutes to complete. 
Below, we enclose the full survey:

\begin{enumerate}
    \item Do you use AI for any part of your professional work? [Yes/No]
    
    \item Please elaborate your response further. Why and what do you use AI for? (Or why not?) [Free text response]

    \item  Among the ways that you use AI, would you disclose any of them to your employers?
    \begin{itemize}
        \item Yes, I would proactively disclose to my employers. 
        \item Yes, I would disclose if my employers ask about it. 
        \item No, I would not disclose at all. 
    \end{itemize}

    \item If you answer "Yes" to the question above, please specify what and how you would disclose to your clients. [Free text response]

    \item Do you have clients or employers explicitly stating their "AI policies" in their job descriptions? (e.g., statements about whether and how employees can use AI for work) [Yes/No]

    \item If you answer "Yes" to the question above, please copy or summarize the statement(s) that you saw from clients and/or employers. [Free text response]

    \item How do you interpret the clients' expectation for AI use based on the "AI policy" that you mentioned above?
    \begin{itemize}
        \item I cannot use AI for any part of my work. 
        \item I can use AI but only to support minor parts of my work, such as: [Free text response]
        \item I can use AI to support key job functions for my work, such as: [Free text response]
        \item I can use AI as much as possible to facilitate my work. 
        \item I can use AI to automate my work. 
    \end{itemize}

    \item In reality, how would you use AI in response to the "AI policy" that you mentioned above?
    \begin{itemize}
        \item I would not use AI for any part of my work. 
        \item I would use AI to support minor parts of my work, such as: [Free text response]
        \item I would use AI to support key job functions for my work, such as: [Free text response]
        \item I would use AI as much as possible to facilitate my work. 
        \item I would use AI to automate my work. 
    \end{itemize}

    \item Consider all the "AI policies" that you have encountered so far, are you able to tell clients and/or employers expectations for using AI at work?
    \begin{itemize}
        \item None of them (around 0\%)
        \item A few of them (around 25\%) 
        \item About half of them (around 50\%)
        \item Many of them (around 75\%)
        \item Most of them (around 100\%)
    \end{itemize}

    \item Do you think employers can tell whether you use AI for work?
    \begin{itemize}
        \item Never (around 0\% of time)
        \item Rarely (around 25\% of time)
        \item Sometimes (around 50\% of time)
        \item Often (around 75\% of time)
        \item Always (around 100\% of time)
    \end{itemize} 

\end{enumerate}

\subsubsection{Survey Participants}
We adopted the same recruitment and screening strategies and expanded to recruit a larger sample size $N=100$ for the survey study.
We report participants' demographics and freelance experiences in \Cref{app:demographics}.


\subsection{Study 2: Survey with Clients on Perspectives and Policies for AI Use and Disclosure}

\subsubsection{Questionnaire Development}
In Study~2, we mirrored the worker survey from Study~1.2 by asking clients on freelance platforms about their expectations for workers’ AI use. Specifically, we invited them to share any ``AI policies'' they had in place and to elaborate on how they expected workers to align their AI use with these guidelines. 
Below, we enclose the full questionnaire from Study~2: 

\begin{enumerate}
    \item Do you allow workers to use AI for work? [Yes/No]
    \item Please elaborate on your response further. Why or what not? [Free text response]
    \item What are the forms of AI use that you allow (e.g., editing text, drafting emails, generating ideas)? Please specify and list out as many as possible. [Free text response]
    \item In what cases would you encourage workers to use AI for work? [Free text response]
    \item In what cases would you be concerned about workers using AI for work? [Free text response]
    \item Based on your own estimate, do you think your freelance workers use AI for work?
    \begin{itemize}
        \item None of them (around 0\%)
        \item A few of them (around 25\%) 
        \item About half of them (around 50\%)
        \item Many of them (around 75\%)
        \item Most of them (around 100\%)
    \end{itemize}
    \item Based on your own estimate, do you think your freelance workers disclose their use of AI for work?
    \begin{itemize}
        \item None of them (around 0\%)
        \item A few of them (around 25\%) 
        \item About half of them (around 50\%)
        \item Many of them (around 75\%)
        \item Most of them (around 100\%)
    \end{itemize}
    \item Based on your own estimate, can you tell whether your freelance workers use AI for work?
    \begin{itemize}
        \item Never (around 0\% of time)
        \item Rarely (around 25\% of time)
        \item Sometimes (around 50\% of time)
        \item Often (around 75\% of time)
        \item Always (around 100\% of time)
    \end{itemize}
    \item Whether and how should workers disclose their use of AI for work?
    \begin{itemize}
        \item Yes, they have to proactively disclose their use of AI.
        \item Yes, they have to disclose their use of AI if I ask.
        \item No, I don't mind whether they disclose their use of AI.
        \item Not applicable. My workers don't use AI at all.   
    \end{itemize}
    \item Please further describe how you ask your workers to disclose AI use, if at all. What information do you ask them to provide? [Free text response]
    \item Do you explicitly state an ``AI policy'' (i.e., statements about whether and how workers can use AI for work) in your job descriptions? [Yes/No]
    \item Please copy your AI policy here. If you don't currently have an AI policy, please write one that reflects how you would like your workers to use AI. [Free text response]
    \item Based on this "AI policy," how do you expect workers to use AI for work, if at all?
    \begin{itemize}
        \item Workers cannot use AI for any part of their work.   
        \item Workers can use AI, but only to support minor parts of their work, such as [Free text response].
        \item Workers can use AI to support primary, key job functions for their work, such as [Free text response].
        \item Workers can use AI as much as possible to facilitate their work.   
        \item Workers can use AI to fully automate their work.  
    \end{itemize}
\end{enumerate}

\subsubsection{Participants}
We adopted the same recruitment and screening strategies and expanded to recruit a larger sample size $N=145$ for the survey study. We report their demographics and freelance experiences in \Cref{app:demographics}.

\subsection{Study 3: Survey with Workers on Interpretation of AI Policies}
In the previous study, clients shared their AI policies and indicated the intended scope of AI use (i.e., No AI use, Use AI for minor tasks, Use AI for major tasks, Use AI as much as possible, or Use AI to fully automate tasks). 
We grouped these policies into the five corresponding categories based on clients' own responses and adopted them as probes for Study 3: Specifically, in this follow-up study, workers evaluated these policies through a survey. 
Each worker was randomly assigned to rate five policies, one from each of the five categories.

\subsubsection{Questionnaire Development}
Besides informed consent and demographic questions, the Study~3 survey consisted of two main components:
\textbf{\textit{Workers’ attitudes toward AI use and disclosure.}}
Building on the same measures used in Study~1.1 and Study~2, we asked workers about their practices and perspectives related to AI in freelance work. These questions addressed: (1) their own adoption and use of AI, (2) their perceptions of clients’ expectations for AI use, and (3) prior experiences with clients’ AI policies and how they interpreted them.
\textbf{\textit{Workers’ responses to clients’ AI policies.}}
We then presented workers with client-defined AI policies (collected in Study~2) and asked them to evaluate each. For every policy, workers rated: (1) how they interpreted the client’s expectations for AI use and disclosure, (2) how they would actually use and disclose AI under that policy, and (3) whether specific AI use cases (these are permitted use cases listed by clients in Study~2) were permitted.
Below, we enclose the full questionnaire from Study~3:

\textit{[Read AI Policy] Below is an "AI policy" that outlines an employer's expectations and guidelines for AI use in the workplace. Please read through it carefully and answer the following questions.}
\begin{enumerate}
    \item Based on this "AI policy," how do you interpret the employer's expectation for AI use for work?
    \begin{itemize}
        \item 	Workers cannot use AI for any part of their work
        \item 	Workers can use AI, but only to support minor parts of their work
        \item 	Workers can use AI to support primary, key job functions for their work
        \item 	Workers can use AI as much as possible to facilitate their work
        \item 	Workers can use AI to fully automate their work
    \end{itemize}
    \item Given this "AI policy," how might you use AI for work in reality?
    \begin{itemize}
        \item 	I would not use AI for any part of my work
        \item 	I would use AI, but only to support minor parts of my work
        \item 	I would use AI to support primary, key job functions for my work
        \item 	I would use AI as much as possible to facilitate my work. 
        \item 	I would AI to fully automate my work. 
    \end{itemize}
    \item Based on this "AI policy," how do you interpret the client's expectation for disclosure of AI use in the workplace?
    \begin{itemize}
        \item 	Workers should proactively disclose their AI use. 
        \item 	Workers should disclose their use of AI when the client asks about it. 
        \item 	Workers do not need to disclose their use of AI. 
        \item 	Not applicable. Workers should not use AI at all. 
    \end{itemize}
    \item Given this "AI policy," how might you disclose AI use in reality?
    \begin{itemize}
        \item 	I would proactively disclose my AI use. 
        \item 	I would disclose my AI use when the client asks about it. 
        \item 	I would not disclose my AI use. 
        \item 	Not applicable. I would not use AI at all. 
    \end{itemize}
    \item Based on this "AI policy," are the following forms of AI use permitted in the workplace? 
    \begin{itemize}
        \item (Read AI use cases)
        \item Rate AI use given the AI policy by: (a) Allowed; (b) Not allowed; (c) It depends; (d) Need manager's approval;	(e) I'm not sure.
    \end{itemize}
\end{enumerate}

\subsubsection{Participants}
Following the same recruitment approach from previous studies, we recruited $N=100$ freelance workers to participate in Study~3.

\section{Participants' Demographics}
\label{app:demographics}
\begin{table}[h!]
    \centering
    \resizebox{\linewidth}{!}{
    \begin{tabular}{p{0.45\linewidth}|r|r|r|r}
        \toprule
         \textbf{Demographics} & \textbf{Study~1.1} & \textbf{Study~1.2} & \textbf{Study~2} & \textbf{Study~3} \\
         \hline
         
         \textbf{Gender} & & & & \\
         Male & 46.3\% & 53\% & 51.38\% & 50.00\% \\
         Female & 46.3\% & 39\% & 33.33\% & 48.96\% \\
         Non-binary/third gender & 4.9\% & 1\% & 1.39\% & 1.04\% \\
         Prefer not to disclose & 2.4\% & 7\% & 2.78\% & 0\% \\
         \hline

         \textbf{Age} & 
         40.35$\pm$11.29 & 36.4$\pm$9.8 & 42.31$\pm$11.20 & 37.29$\pm$9.88 \\
         \hline

         \textbf{Ethnicity} & & & & \\
         White/Caucasian & 61\% & 57.1\% & 61.11\% & 48.41\% \\
         African/Black American & 14.6\% & 21.4\% & 9.72\% & 23.81\% \\
         Asian & 4.9\% & 8.2\% & 6.94\% & 7.14\% \\
         Hispanic, Latino, Latinx & 4.9\% & 6.1\% & 4.17\% & 3.17\% \\
         Multi-racial or multi-cultural & 7.3\% & 6.1\% & 12.50\% & 15.87\% \\
         Native American & 0\% & 1\% & 0\% & 0\% \\
         Prefer not to disclose & 7.3\% & 0\% & 2.78\% & 0.79\% \\
        \bottomrule
    \end{tabular}}
\end{table}

\section{Freelance Service Offered/Sought by Workers and Clients}
\label{app:freelance-type}
\textit{Note: }Each worker can provide more than one type of service.
\begin{table}[H]
    \centering
    \resizebox{\linewidth}{!}{
    \begin{tabular}{p{0.45\linewidth}|r|r|r|r}
    \toprule
         \textbf{Freelance Service Type} & \textbf{Study~1.1} & \textbf{Study~1.2} & \textbf{Study~2} & \textbf{Study~3} \\
         \hline
         Healthcare and Medicine	
         & 12\% &	7\% &	16\% &	14\% \\
         \arrayrulecolor{gray!40}\hline
         
         Information Technology (IT) and Engineering	
         & 23\% &	27\% &	23\% &	22\% \\
         \arrayrulecolor{gray!40}\hline
         
         Education and Training	
         & 8\% &	5\% &	6\% &	8\% \\
         \arrayrulecolor{gray!40}\hline
         
         Sales and Marketing	
         & 15\% &	12\% &	13\% &	14\% \\
         \arrayrulecolor{gray!40}\hline
         
         Finance and Accounting	
         & 5\% &	18\% &	13\% &	10\% \\
         \arrayrulecolor{gray!40}\hline
         
         Human Resources (HR) and Recruitment	
         & 15\% &	2\% &	16\% &	1\% \\
         \arrayrulecolor{gray!40}\hline
         
         Administration and Support	
         & 13\% &	3\% &	11\% &	7\% \\
         \arrayrulecolor{gray!40}\hline
         
         Manufacturing and Production	
         & 5\% &	3\% &	9\% &	4\% \\
         \arrayrulecolor{gray!40}\hline
         
         Legal Services	
         & 5\% &	1\% &	2\% &	3\% \\
         \arrayrulecolor{gray!40}\hline
         
         Hospitality and Tourism	
         & 7\% &	7\% &	9\% &	18\% \\
         \arrayrulecolor{gray!40}\hline
         
         Creative Arts and Media	
         & 6\% &	2\% &	8\% &	3\% \\
         \arrayrulecolor{gray!40}\hline
         
         Real Estate and Property Management	
         & 2\% &	4\% &	3\% &	4\% \\
         \arrayrulecolor{gray!40}\hline
         
         Transportation and Logistics	
         & 2\% &	1\% &	2\% &	1\% \\
         \arrayrulecolor{gray!40}\hline
         
         Science and Research	
         & 3\% &	7\% &	3\% &	1\% \\
         \arrayrulecolor{gray!40}\hline
         
         Public Sector and Government	
         & 12\% &	5\% &	6\% &	1\% \\
         \arrayrulecolor{gray!40}\hline
         
         Other	
         & 17\% &	17\% & 	14\% &	9\% \\
    \arrayrulecolor{black}\bottomrule
    \end{tabular}}
    \label{tab:s1-1-freelance-type}
\end{table}

\end{document}